\documentclass[vecphys]{svmult}

\usepackage{makeidx}         
\usepackage{graphicx}        
\usepackage{multicol}        
\usepackage[bottom]{footmisc}

\makeindex             

\def\la{\mathrel{\hbox to 0pt{\lower 3.5pt\hbox{$\mathchar"218$}\hss}
      \raise 1.5pt\hbox{$\mathchar"13C$}}}
\def\ga{\mathrel{\hbox to 0pt{\lower 3.5pt\hbox{$\mathchar"218$}\hss}
      \raise 1.5pt\hbox{$\mathchar"13E$}}}

\begin{document}

\title*{Gaussian analysis of the CMB with the smooth tests of goodness
  of fit}
\titlerunning{Smooth tests of goodness of fit}
\author{R.B. Barreiro\inst{1} \and J.A. Rubi\~no-Mart\'\i n\inst{2} \and
  E. Mart\'\i nez-Gonz\'alez\inst{1}}
\institute{Instituto de F\'\i sica de Cantabria, CSIC -- Universidad de
Cantabria, Avda. de los Castros s/n, 39005, Santander, Spain \\
\texttt{barreiro@ifca.unican.es, martinez@ifca.unican.es} \and
Instituto de Astrof\'\i sica de Canarias, C/ V\'{\i}a Lactea s/n,
38200, La Laguna, Spain \\ \texttt{jalberto@iac.es}
}
%
\maketitle

The study of the Gaussianity of the cosmic microwave background (CMB)
radiation is a key topic to understand the process of structure
formation in the Universe. In this paper, we review a very useful tool
to perform this type of analysis, the Rayner \& Best smooth
tests of goodness of fit. We describe how the method has
been adapted for its application to imaging and interferometric
observations of the CMB and comment on some recent and future
applications of this technique to CMB data.

\section{Introduction}
\label{sec:intro}

The study of the Gaussianity of the cosmic microwave background (CMB)
fluctuations has become a very useful tool in constraining theories of
structure formation. The standard inflationary scenario predicts
Gaussian fluctuations whereas other competitive theories would imprint
non-Gaussian signatures on the CMB (see \cite{bb:bar04} for a review).
Therefore, the study of the Gaussianity of the CMB can help to discard
or constrain some of these theories.  Moreover, secondary effects
(e.g. gravitational lensing, Rees-Sciama effect, Sunyaev-Zeldovich
effect...), astrophysical emissions and systematics may as well leave
non-Gaussian imprints on the CMB, which should not be confused with
intrinsic non-Gaussianity.

Given the importance of this type of analysis and taking into account
that different methods may be sensitive to different kinds of
non-Gaussianity, many tools have been developed for the study of the
temperature distribution of the CMB. Among others, they include the
Minkowski functionals \cite{bb:kom03}, the bispectrum \cite{bb:fer98},
wavelet techniques \cite{bb:bar02}, geometrical estimators
\cite{bb:mon06} or smooth tests of goodness of fit \cite{bb:ali05}.

The interest for this type of analysis has increased even more since
the release of the WMAP data \cite{bb:ben03}. A large number of
different techniques have been applied to study whether these data
follow or not a homogeneous and isotropic Gaussian random field,
finding in some cases unexpected results. In particular, a significant
number of works have reported deviations from Gaussianity and/or
isotropy, whose origin is uncertain
(e.g. \cite{bb:vie04,bb:eri04,bb:han04,bb:cay05,bb:cop06,bb:cru06,
bb:mce06,bb:wia06}, see also \cite{bb:mar06} for a review).

In this paper, we review the Rayner and Best smooth tests of goodness
of fit for the study of the Gaussianity of the CMB. In section
\ref{sec:gof} we describe the test and how to adapt the method for its
application to CMB observations. A discussion about current and future
applications to different CMB datasets is given in section
\ref{sec:applications}. Finally our conclusions are summarised in
section \ref{sec:conclusions}.

\section{The Rayner and Best smooth tests of goodness of fit}
\label{sec:gof}

Given a statistical variable $X$ and n independent realizations
${x_i}$, $i=1,...,n$, we want to test if $X$ follows a given
probability density function (pdf) $f(x)$. The
smooth tests of goodness of fit (gof) allows one to discriminate
between a predetermined pdf $f(x)$ (null hypothesis) and a second
one that deviates smoothly from the former (alternative hypothesis).

Among the possible forms for the alternative pdf, Rayner \& Best
\cite{bb:ray89,bb:ray90} consider:
\begin{equation}
f_k(x,\theta)=C(\theta)\exp \left[\sum_{i=1}^{k}\theta_i h_i(x)
\right]f(x) \ ,
\end{equation}
where $\theta=(\theta_1,...,\theta_k)$ is a set of $k$ parameters that
allows for smooth deviations of the alternative hypothesis with
respect to $f(x)$, $C(\theta)$ is a normalisation constant that
ensures that $f_k$ is normalised to 1 and $h_i$ form a complete set
of orthonormal functions of $f$. Note that for $\theta=0$ we recover
$f(x)$, therefore, our statistical analysis consists on testing the
null hypothesis $H_0 : \{\theta=0\}$ versus the alternative hypothesis
$H_1 : \{\theta \ne 0\}$.

To perform this analysis, the score statistic is used. This is a
quantity which is closely related to the likelihood ratio (see
e.g. \cite{bb:ray89}). For the Rayner \& Best smooth tests of gof, the score
statistic associated to the $k$ alternative is given by
\begin{eqnarray}
\label{eq:ui}
S_k & = &\sum_{i=1}^k U_i^2 \\
{\rm with} \ \ U_i & = \frac{1}{\sqrt n}&\sum_{j=1}^n h_i(x_j) 
\end{eqnarray}
Large values of $S_k$ (or of $U_i^2$) reject the null hypothesis.

In the case of testing if our data follow a Gaussian distribution of
zero mean and unit dispersion, the $h_i$ are given by the (normalised)
Hermite Chebishev polynomials.\footnote{The form of the $h_i$ for
other usual distributions (e.g. uniform, exponential) can be found in
\cite{bb:ray89}.} In this case, it is possible to write the $U_i^2$
quantities in terms of the moments of order $k$, $\mu_k$, of the
data\footnote{The moment of order $k$ of the data is defined as
$\mu_k=\sum_{j=1}^n y_j^k/n$}:
\begin{eqnarray}
\label{eq:ui2}
U_1^2&=&n \mu_1^2 ~~~~~~~~~~~~~~~~~ U_3^2=\frac{n}{6}(\mu_3-3\mu_1)^2
\nonumber \\  
U_2^2&=&\frac{n}{2}(\mu_2-1)^2 ~~~~~~~
U_4^2=\frac{n}{24}(\mu_4-6\mu_2+3)^2 
\end{eqnarray}
If the Gaussian hypothesis holds, the $U_i^2$ follow a $\chi^2_1$
distribution when $n \to \infty$. This allows one to determine easily
the significance of any possible deviation from Gaussianity by
comparing the value of the $U_i^2$ of the data with a $\chi^2_1$.

We must point out that the proposed technique is designed to test if
the data follow a univariate Gaussian. Thus, for optimality, it should
be applied to independent data. However, the CMB signal is correlated
at all scales and the noise may as well present
correlations. Therefore, before applying the gof test, it is necessary
to transform the data to make them as independent as possible. 

One possibility is to obtain the Cholesky decomposition of the
correlation matrix of the data (including signal plus noise) $C=LL^t$
and then multiply the ${x_i}$ by the inverse of the Cholesky matrix,
i.e. $y_i=\sum_{j}L_{ij}^{-1}x_j$. The constructed $y_i$ are
uncorrelated, have zero mean and unit dispersion. Moreover if the data
are Gaussian, they also follow a normal distribution and are
independent.  This decorrelation technique has been used for analysing
the MAXIMA data with different smooth tests of gof
\cite{bb:cay03,bb:ali03a,bb:ali03b}. Nevertheless, the preprocessing
of the data has been improved in subsequent works through the use of a
signal-to-noise decomposition, which is explained in the next
subsection.

\subsection{Signal-to-noise decomposition}

The signal to noise decomposition was introduced in the CMB field by
\cite{bb:bon95}, whereas \cite{bb:ali05} applied this formalism
jointly with the gof test. This technique allows one to construct
uncorrelated eigenmodes from the data which are also associated to a
certain signal-to-noise ratio.

Let us consider a set of CMB data $d_i$, $i=1,...,n$, where $i$
corresponds to a given position in the sky. This can be written as
\begin{equation}
d_i=s_i+n_i
\end{equation}
where $s_i$ and $n_i$ are the contributions from the CMB signal and
noise, respectively. The mean values of signal and noise are assumed
to be zero and their correlation matrices are given by $S_{ij}=\left<s_i
s_j \right>$ and $N_{ij}=\left<n_i n_j \right>$ where the brackets
indicate average over many realizations.

The signal-to-noise eigenmodes are defined as
\begin{equation}
\vec{\xi} = R_A^tL_N^{-1} \vec{d} \, \label{eigenmode}
\end{equation}
where $L_N$ is the Cholesky matrix of $N$, i.e. $N=L_N L_N^t$, and $R$
is the rotation matrix that diagonalizes the matrix
$A=L_{N}^{-1}SL_{N}^{-t}$. The eigenvalues of this diagonalization are
denoted by $E_i$. Let us now construct the quantities $y_i$:
\begin{equation}
y_i=\frac{\xi_i}{\sqrt{1+E_i}} \label{yi}
\end{equation}
It can be shown that these quantities are uncorrelated and have zero
mean and unit dispersion. Moreover, if the data $\vec{d}$ are
multinormal, then the $y_i$ are distributed according to a Gaussian
pdf, since all the applied transformations are linear. In this case
the $y_i$ are also independent. Therefore we are in the optimal conditions
to apply the gof tests to the quantities $y_i$.

In addition, we also have information about the signal-to-noise ratio
of the $i$ eigenmode, which is given by $\sqrt{E_i}$. This means that
eigenmodes with low values of $E_i$ are dominated by noise and may be
discarded from the analysis. Therefore, in practice, the gof test will
be applied to the subset of $y_i$ such that its signal-to-noise ratio
is greater than a given threshold, i.e. $E_i > E_{cut}$.  Thus, this
decomposition allows us not only to obtain uncorrelated variables but
also to select the fraction of the data where the signal contribution
dominates over the noise.

\subsection{Application to interferometer observations}

The previous technique has been adapted to deal with interferometric
data by \cite{bb:ali05} and applied to VSA data in \cite{bb:rub06}.

Let us consider an interferometer observing a small region of the
sky at frequency $\nu$, for which the flat-sky approximation is
valid. In this case the complex visibility, which is the response of
the interferometer at the considered frequency, is given by
\begin{equation}
V(\vec{u},\nu)= \int{P(\hat{\vec{x}},\nu}) B(\hat{\vec{x}},\nu)
\exp(i2\pi\vec{u}\hat{\vec{x}})d\hat{\vec{x}}  
\end{equation}
where $\hat{\vec{x}}$ corresponds to the angular position of the
observed point on the sky and $\vec{u}$ is the baseline vector in
units of the wavelength of the observed
radiation. $P(\hat{\vec{x}},\nu)$ is the primary beam of the antennas
(normalized to unity at its peak) and $B(\hat{\vec{x}},\nu)$
corresponds to the brightness distribution on the sky.

Of course, for a realistic instrument, the effect of instrumental
noise should be also taken into account. Therefore, the $i$th baseline
$\vec{u}_i$ of the interferometer will measure
\begin{equation}
d(\vec{u}_i,\nu)=V(\vec{u}_i,\nu)+n(\vec{u}_i,\nu)
\end{equation}
where $n(\vec{u}_i,\nu)$ corresponds to the instrumental noise of the
$\vec{u}_i$ visibility.  

Let be $N$ the total number of complex visibilities observed by the
interferometer. Since the measured quantities are complex, the number
of elements that constitute the data are $N_d=2N$, corresponding to
the real and imaginary parts of each observed visibility.

Testing the Gaussianity of the measured visibilities is equivalent to
testing the joint Gaussianity of their real and imaginary
parts. Therefore the signal-to-noise decomposition can be applied
directly to these quantities (so we will have a total of $N_d$
eigenmodes). The correlation matrix $S$ of the real and imaginary
parts of $V(\vec{u}_i,\nu)$ (i.e. the correlation matrix of the
signal) can be computed following the work of \cite{bb:hob02} whereas
the noise correlation matrix is determined by the characteristics of
the instrument. Once the signal-to-noise eigenmodes have been
obtained, the gof technique can be applied to test the Gaussianity of
these quantities (or of a subset of them with the highest
signal-to-noise ratio).  As in the previous case, if the data are
distributed as a multinormal, the constructed eigenmodes are
independent and follow a Gaussian distribution of zero mean and unit
dispersion.

A complementary analysis can also be performed on the phases of the
decorrelated visibilities. If the data are Gaussian, the phases should
follow a uniform distribution. This can be tested using the Rayner \&
Best smooth tests of gof by considering the appropriate $h_i$ in
equation (\ref{eq:ui}) (see \cite{bb:ray89,bb:ali05} for
details). However, \cite{bb:ali05} found that, for their considered
examples, the phase analysis was less sensitive to deviations from
Gaussianity than the test based on the real and imaginary parts of the
visibilities.

\subsection{Some comments about the method}
One of the advantages of the Gaussianity analysis based on the gof
test and the signal-to-noise formalism is that it is well suited for
the study of many different kinds of CMB observations. In particular,
it can be adapted to deal with most of the problematics found in real
data. For instance, it is not affected by the presence of holes in the
data or by the use of irregular masks and it can easily deal with
anisotropic and/or correlated noise. Also, as already explained, it
can be applied to imaging or interferometric data. Another interesting
feature of the method is that it allows one to choose that fraction of
the data with a signal-to-noise ratio above a certain threshold. In
addition, as will be discussed in the next section, it is a very
sensitive technique, being able to detect different type of deviations
in the data (such as intrinsic non-Gaussianity, systematic effects or
anisotropy of the local power spectrum).

The main shortcoming of the technique is the large amount of CPU
required to calculate the signal-to-noise eigenmodes, since it
involves the diagonalization of large matrices (of size $n\times n$,
where $n$ is the number of data to be analysed). However, the method
uses only a fraction of the eigenmodes (those whose signal-to-noise
ratio is higher than a given threshold) and therefore it is not
necessary to obtain all the eigenmodes and eigenvalues of the
problem. To take advantage of this fact, \cite{bb:rub06}
proposes the use of the Arnoldi algorithm which significantly speeds
the calculation of the required $y_i$. This method is based on the
construction of a matrix $H$ of dimension $m \times m$ (with $m < n$)
such that it is possible to construct a good approximation to certain
eigenvectors and eigenvalues of $A$ from those of $H$. In particular,
the eigenvectors that are well approximated correspond to those with
higher eigenvalues. From these quantities it is also possible to
construct those eigenmodes with higher signal-to-noise ratio, i.e.,
those that are kept for the analysis (see \cite{bb:rub06,bb:saa92} for
details). This means that we have significantly reduced the
computational cost of the analysis, since we are working with
a matrix of size $m \times m$ instead of $n \times n$.

\section{Applications to CMB data}
\label{sec:applications}
The gof tests were firstly introduced in the CMB field by
\cite{bb:cay03}, which carried out a Gaussianity analysis of the
MAXIMA data \cite{bb:han00}. The results showed that the data were
compatible with Gaussianity (see also \cite{bb:ali03a,bb:ali03b}).

A more recent application of the Rayner \& Best gof test has been
carried out by \cite{bb:rub06}, that present a Gaussianity analysis of
the Very Small Array (VSA) data \cite{bb:tay03,bb:gra03,bb:dic04}. The
VSA is an interferometer sited at the Teide Observatory (Tenerife)
designed to observe the sky on scales going from 2$^\circ$ to 10$'$
and operates at frequencies between 26 and 36 GHz (see \cite{bb:wat03}
for a detailed description).

In the analysis, most of the fields observed by the VSA were found to
be compatible with Gaussianity. However, deviations from Gaussianity
were detected in the $U_2^2$ statistic in three cases. After a
thorough analysis of the possible origins of these detections, the
authors concluded that one of the deviations was associated to a
residual systematic effect of a few visibility points, which, when
corrected, have a negligible effect on the angular power spectrum. A
second detection seemed to have its origin in a deviation of the local
power spectrum of the considered field with respect to the power
spectrum estimated from the complete dataset. This deviation was
found at angular scales around the third angular peak
($\ell=700-900$). If the affected visibilities were removed, a
cosmological analysis based only on this modified power spectrum and
the COBE data showed no differences except for the physical baryon density,
which decreased by 10 per cent and got closer to the value obtained
from Big Bang Nucleosynthesis. Finally, the third deviation from
Gaussianity was found in observations of the Corona Borealis
supercluster region \cite{bb:gen05}. In this case, the non-Gaussianity
was identified as intrinsic to the data, probably due, at least in
part, to the presence of Sunyaev-Zeldovich emission in the
region. This result has been later confirmed with the measurements of
the MITO telescope in this region \cite{bb:bat06}.  A combined maximum
likelihood analysis of the MITO and the VSA data provided a weak
detection of a faint signal compatible with a SZ effect, characterized
by a Comptonization parameter of $y=(7.8^{+5.3}_{-4.4}) \times
10^{-6}$, at 68\% CL.

An application of the gof technique to the Archeops data is currently
ongoing \cite{bb:cur06}. Archeops is a balloon-borne experiment, which
is dedicated to measure the CMB temperature anisotropies from large
to small angular scales \cite{bb:ben02,bb:beno03}. It has also been
designed as a test bed for the forthcoming Planck high frequency
instrument. The preliminary results show the good performance of the
method, that is able to deal with the presence of anisotropic and
correlated noise in the data.

The application of the gof technique to the WMAP data \cite{bb:ben03}
is of great interest and is currently in progress. Due to the large
amount of data observed by this experiment, a whole sky analysis at
full resolution is unfeasible, due to the large computational
resources required for the signal-to-noise decomposition. However, two
types of complementary tests are possible: an analysis of the full-sky
at low-resolution and a study of small regions of the sky at high
resolution. Given the sensitivity of the gof tests to detect
deviations from a homogeneous and isotropic Gaussian random field,
this analysis could shed new light on some of the anomalies reported
for the WMAP data.

\section{Conclusions}
\label{sec:conclusions}
We have reviewed the Rayner \& Best smooth tests of goodness of fit
and its applications to CMB data. One of the most interesting features
of this method is that it can deal with most of the problematics found in
real data such as the use of irregular masks or the presence of
anisotropic and/or correlated noise. In addition, it has been adapted
to deal either with imaging or interferometric observations. The main
shortcoming of the technique is the large computational cost required
to perform the signal-to-noise decomposition of the data. However,
this problem can be significantly alleviated by the use of approximate
methods such as the Arnoldi algorithm.

The recent and current applications of the gof tests to different
datasets are showing its good performance. Most notably, the method
has been able to detect deviations from a homogeneous and isotropic
Gaussian field in the VSA data, which were associated to very
different origins: residual systematics, a deviation of the local
power spectrum with respect to the global one and non-Gaussianity
intrinsic to the data. It is important to mention that Gaussianity
analyses had already been performed in the VSA dataset using other
methods \cite{bb:sav04,bb:smi04} but neither the residual systematics
nor this small deviation of the power spectrum were
detected. Therefore we believe that this method constitutes a very
useful tool for the statistical analysis of CMB data.

\section*{Acknowledgements}
RBB and EMG thank financial support from the Spanish MEC project
ESP2004-07067-C03-01

%

\begin{thebibliography}{99.}

\bibitem{bb:ali03a} A.M. Aliaga, E. Mart\'\i nez-Gonz\'alez,
L. Cay\'on, F. Arg\"ueso, J.L. Sanz, R.B. Barreiro: 
New Ast. Rev., \textbf{47}, 821 (2003)

\bibitem{bb:ali03b} A.M. Aliaga
et al.:
New Ast. Rev., \textbf{47}, 907 (2003)

\bibitem{bb:ali05} A.M. Aliaga, J.A. Rubi\~no-Mart\'\i n, E. Mart\'\i
nez-Gonz\'alez, R.B. Barreiro, J.L. Sanz: MNRAS, \textbf{356}, 1559 (2005)

\bibitem{bb:bar02} R.B. Barreiro, M.P. Hobson: MNRAS, \textbf{327},
813 (2002)

\bibitem{bb:bar04} N. Bartolo, E. Komatsu, S. Matarrese, A. Riotto:
Phys. Rep., \textbf{402}, 103 (2004)

\bibitem{bb:bat06} E.S. Battistelli et al.: ApJ, \textbf{645}, 826 (2006)

\bibitem{bb:ben03} C.L. Bennett et al.: ApJS, \textbf{148}, 1 (2003)

\bibitem{bb:ben02} A. Beno\^\i t et al.: Astropart. Phys.,
\textbf{17}, 101 (2002)

\bibitem{bb:beno03} A. Beno\^\i t et al.: A\&A, \textbf{399}, L19 (2003)

\bibitem{bb:bon95} J.R. Bond: Phys. Rev. Lett., \textbf{74}, 4369 (1995)

\bibitem{bb:cay03} L. Cay\'on, F. Arg\"ueso,  E. Mart\'\i nez-Gonz\'alez,
J.L. Sanz: MNRAS, \textbf{344}, 917 (2003)

\bibitem{bb:cay05} L. Cay\'on, J. Jin, A. Treaster: MNRAS,
\textbf{362}, 826 (2005)

\bibitem{bb:cop06} C.J. Copi, D. Huterer, D.J. Schwarz, G.D. Starkman:
MNRAS, \textbf{367}, 79 (2006)

\bibitem{bb:cur06} A. Curto, J. Aumont, J.F. Mac\'\i as-P\'erez,
E. Mart\'\i nez-Gonz\'alez, R.B. Barreiro, D. Santos: in preparation 

\bibitem{bb:cru06} M. Cruz, M. Tucci, E. Mart\'\i nez-Gonz\'alez,
P. Vielva: MNRAS, \textbf{369}, 57 (2006)

\bibitem{bb:dic04} C. Dickinson et al.: MNRAS, \textbf{353}, 732 (2004)

\bibitem{bb:eri04} H.K. Eriksen, F.K. Hansen, A.J. Banday,
K.M. G\'orski, P.B. Lilje: ApJ, \textbf{605}, 14 (2004)

\bibitem{bb:fer98} P.G. Ferreira, J. Magueijo, K.M. G\'orski: ApJ,
\textbf{503}, L1 (1998)

\bibitem{bb:han00} S. Hanany et al.: ApJ, \textbf{545}, L5 (2000)

\bibitem{bb:han04} F.K. Hansen, A.J. Banday, K.M. G\'orski: MNRAS,
\textbf{354}, 641 (2004)

\bibitem{bb:hob02} M.P. Hobson, K. Maisinger: MNRAS, \textbf{334}, 569
(2002)

\bibitem{bb:gen05} R. G\'enova-Santos et al.: MNRAS, \textbf{363}, 79
(2005) 

\bibitem{bb:gra03} K. Grainge et al.: MNRAS, \textbf{341}, L23 (2003)

\bibitem{bb:kom03} E. Komatsu et al.: ApJS, \textbf{148}, 119 (2003)

\bibitem{bb:mar06} E. Mart\'\i nez-Gonz\'alez: Cosmic Microwave
Background anisotropies: the power spectrum and beyond. In:
\textit{Data Analysis in Cosmology}, ed by V. Mart\'\i nez,
E. Mart\'\i nez-Gonz\'alez, M.J. Pons-Border\'\i a, E. Saar
(Springer-Verlag), in press


\bibitem{bb:mce06} J.D. McEwen, M.P. Hobson, A.N. Lasenby, D.J. Mortlock: 
MNRAS, \textbf{371}, L50 (2006)

\bibitem{bb:mon06} C. Monteser\'\i n, R.B. Barreiro, E. Mart\'\i
nez-Gonz\'alez, J.L. Sanz: MNRAS, \textbf{371}, 312 (2006)

\bibitem{bb:ray89} J.C.W. Rayner, D.J. Best: \textit{Smooth Tests of
Goodness of Fit} (Oxford University Press, New York, 1989)

\bibitem{bb:ray90} J.C.W. Rayner, D. J. Best: International Statistical
Rev., \textbf{58}, 9 (1990)

\bibitem{bb:rub06} J.A. Rubi\~no-Mart\'\i n et al.: MNRAS,
\textbf{369}, 909 (2006)

\bibitem{bb:saa92} Y. Saad: \textit{Numerical Methods for Large
Eigenvalue Problems} (Manchester Univ. Press, Manchester, 1992)

\bibitem{bb:sav04} R. Savage et al.: MNRAS, \textbf{349}, 973 (2004)

\bibitem{bb:smi04} S. Smith et al.: MNRAS, \textbf{352}, 887 (2004)

\bibitem{bb:tay03} A.C. Taylor et al.: MNRAS, \textbf{341}, 1066 (2003)

\bibitem{bb:vie04} P. Vielva, E. Mart\'\i nez-Gonz\'alez,
R.B. Barreiro, J.L. Sanz, L. Cay\'on: ApJ, \textbf{609}, 22 (2004)

\bibitem{bb:wat03} R.A. Watson et al: MNRAS, \textbf{341}, 1057 (2003)

\bibitem{bb:wia06} Y. Wiaux, P. Vielva, E. Mart\'\i nez-Gonz\'alez,
P. Vandergheynst: Phys. Rev. Lett. \textbf{96}, 151303 (2006)



\end{thebibliography}
%



\printindex
\end{document}